\documentclass[doublecol]{epl2} 
\usepackage{amssymb}
\newcommand{\titlename}{Explicit solver for the electronic $V$-representation problem}
\newcommand{\bra}[1]{\langle #1\vert}
\newcommand{\ket}[1]{\vert#1\rangle}

\newcommand{\eqref}[1]{Eq.~(\ref{#1})}
\newcommand{\vertiii}[1]{{\left\vert\kern-0.25ex\left\vert\kern-0.25ex\left\vert #1 
    \right\vert\kern-0.25ex\right\vert\kern-0.25ex\right\vert}}

\usepackage{hyperref}
\hypersetup{
  pdftitle={\titlename},
  colorlinks,%
  citecolor=black,%
  filecolor=black,%
  linkcolor=black,%
  urlcolor=black,
  pdfauthor={James Daniel Whitfield},
  pdfsubject={},
}

\graphicspath{{../figures/}}
\title{\titlename}

\author{J. D. Whitfield\inst{1}\thanks{E-mail: \email{jdwhitfield@gmail.com}}}
\shortauthor{J. D. Whitfield}
\institute{                    
  \inst{1} Vienna Center for Quantum Science and Technology\\University of Vienna, Department of Physics, Boltzmanngasse 5, Vienna, Austria 1190
}

\abstract{
One route to numerically propagating quantum systems is time dependent density functional theory (TDDFT). The application of TDDFT to a particular system's time evolution is predicated on $V$-representability which we have analyzed in a previous publication. In this work, we provide new insights concerning lattice $V$-representability using an newly developed explicit solver for the time-dependent Kohn-Sham potential which contrast with implicit solvers studied in the past few years. 
We present and interpret the force-balance equation central to our numerical method, describe details of its implementation, and present illustrative numerical results. A new characterization of $V$-representability for one-electron systems is also included. Taken together, the results here open the door to deeper theoretical and numerical investigations of the foundations of TDDFT.
}

\begin{document}

\maketitle

\section{Introduction}

Important classes of time-evolution algorithms widely employed by chemists and physicists are based on reduced descriptions of the wave function. These include methods focused on the two-body reduced density matrix and the electron density (the diagonal of the one-body reduced density matrix). 

Since all interactions of non-relativistic Hamiltonians are between at most two electrons, the $N$-electron wave function, $\Psi$, contains more information than necessary.  For this reason, the two-electron reduced density matrix (2RDM) contains enough information to characterize properties of non-relativistic quantum systems~\cite{Mazziotti11}.   Unfortunately, one must characterize the set of valid 2RDMs corresponding to a valid $N$-electron wave function; this is known as the \textit{$N$-representability problem}~\cite{Coulson60}.  The $N$-representability problem was proven to be QMA-complete~\cite{Lui07} highlighting the theoretical difficulty of 2RDM methods.  Nonetheless, there has been successful efforts to perform time evolution using 2RDM methods~\cite{Lackner15}.

An even more concise description is afforded by the ground state one-electron probability density, 
\begin{equation}
	n_t=\textrm{diag}\left(\textrm{Tr}_{2\ldots N}\ket{\Psi_t}\bra{\Psi_t}\right)
\label{eq:density}
\end{equation}
which we will simply refer to as the density.  The Hohenberg-Kohn theorems dictate that the ground state density, $n_{\lambda_0}$, is sufficient to characterize all properties of the quantum system~\cite{Hohenberg64}.  This provides the basis for density functional theory (DFT).  While theoretically compelling, many functionals to efficiently compute properties from the density are unknown. Moreover, the universal functional necessary for evaluating the energy is unlikely to be determined numerically even to only polynomially accuracy in the size of the system. Despite the numerous approximations to the universal functional, computational complexity arguments \cite{Schuch09} showed that obtaining the numerically exact functional is intractable even with quantum computation.

The corresponding time-dependent result~\cite{Whitfield14a} states, for sufficiently well-behaved systems, the potential can be computed efficiently with access to a quantum computer. Unlike the ground state result, the time-dependent complexity analysis relied on the Kohn-Sham (KS) construction lying at the heart of nearly all practical schemes for DFT.  In this letter, we return to the analysis begun in our previous work~\cite{Whitfield14a} using a combination of theory and numerics.

First let us define the general $V$-representation problem associated with the KS system as the task of constructing a model system which has the same expectation values on selected observables as a target system. 
$V$-representability refers to the existence of solutions to this problem when different constraints are placed on the model system.  The time-dependent subset of $V$-representation problems considers as input an initial state and the target trajectory of selected observables, and the task is to find the correct time-dependent fields for a specified control Hamiltonian. This general framework is not limited to electronic systems as illustrated by a study of this problem in the context of spin systems  \cite{Tempel12}.  Here, attention will focus on the electronic $V$-representation problem where the tasks is to construct a KS system governed by a time-dependent potential $V(t)$ such that the KS density matches the density of a specified interacting many-electron system at all times. 

For fermionic simulations, Ref.~\cite{Leeuwen99} was first to give a constructive solution to time-dependent $V$-representation problem. This was challenged in Ref.~\cite{Baer08} where counter-examples were presented to this construction. These counter-examples were largely addressed by Refs.~\cite{Li08,Farzanehpour12} through a detailed analysis of densities evolving on lattices.  In separate work, an implicit solution using a fixed point mapping has been formulated directly in the continuum limit~\cite{Ruggenthaler11,Ruggenthaler12,Nielsen13}.  Here, we will present an explicit method based on the algorithm analyzed in \cite{Whitfield14a}. 

The paper begins with the force-balance equation, then turn towards the implementation details of the solver for the $V$-representation problem. We give some numerical examples before discussing single-electron $V$-representability theorems. Finally, an outlook closes the letter.

\section{The force-balance equation}
The non-interacting $V$-representation problem requires that the fictitious system's wave function, $\ket{\Phi_t}$, evolve such that its density expectation value, $\bra{\Phi_t}\hat n\ket{\Phi_t}$, matches a target evolution, $n^{aim}(t)$.  The force balance equation determines the required instantaneous potential to correctly construct the KS system.  Note that forces enter at second order of evolution as anticipated by Newton's law: $F=ma$.

The force-balance equation is easily derived from the second derivative of the density following the  Heisenberg equation~\cite{Whitfield14a,Farzanehpour12}.
If we aim for a target evolution, then we should have that $\partial_t^2n^{aim}$ is equal to
$i\bra{\Phi_t}  [\hat H, \partial_t \hat n]\ket{\Phi_t}$. Expanding this commutator, we have two terms, $i[\hat T,\partial_t \hat n]$ and $i[\hat V,\partial_t \hat n]$ which we will physically interpret as well as given some guidance on numerical implementation.

We first discuss the acceleration which the forces must cause.
The free acceleration, also called the momentum-stress tensor \cite{Leeuwen99}, $\hat Q_x=i[\hat T,\partial_t \hat n_x]=-\langle [\hat T,[\hat T,\hat n_x]]\rangle$ 
is independent of the potential operator.  To evaluate the free acceleration, we use $\hat Q=2\Re [\textrm{diag}\left(T\hat\rho^{(1)} T -\hat\rho^{(1)}T^2 \right)]$ with $\hat \rho^{(1)}_{ij}= a_j^\dag a_i$ as the one-body density matrix. Because we are considering fermions, $a_ia_j^\dag=\delta_{ij}-a_j^\dag a_i$ and $a_ia_j=-a_ja_i$. 
The forced acceleration is given by the difference between the free acceleration and the target acceleration $\hat S=\partial_t^2 n^{aim}-\hat Q$.  The forced acceleration then determines the forces required from the potential.  

The forces enter through the term $i[\hat V,\partial_t \hat n]$ which can be recast into two useful forms; one illustrating a connection to forces and the other geared towards determining the potential. 
The first form we examine is 
\begin{equation}
i[\hat V,\partial_t \hat n_j]=
\sum_k(V_j-V_k) T_{kj}(  a_j^\dag  a_k+ a_k^\dag  a_j ).
\end{equation}
This form gives a nice analogy to the real space forces as $F(x)=-\nabla V(x)$.
We note that $i\bra{\Phi}[\hat V,\partial_t \hat n_j]\ket{\Phi}=2\sum_k(V_j-V_k) \; \Re [T_{kj}^\Phi] $ has the form of a generalized discrete gradient.  Here, the real part of $T^{\Psi}_{ij}= T_{ij}\bra{\Psi}  a_i^\dag  a_j\ket{\Psi}$ includes both the influence of the probability mass at each site as well as the underlying spatial metric.

The second form is more applicable to numerical simulation: 
$i\bra{\Phi}[\hat V,\partial_t \hat n_j]\ket{\Phi}$ $=\sum_r\bra{\Phi}\left(i[\hat n_r,\partial_t \hat n]\right)\ket{\Phi} V_r$.
To introduce a simpler expression, it will be advantageous to define $\mathcal{M}(A)_{ij} = A_{ij}-\delta_{ij}\left( \sum_k A_{jk}\right)$.  Then, the force-balance equation can be expressed as:
\begin{equation}
i\bra{\Phi}[\hat V,\partial_t \hat n_r]\ket{\Phi}=\sum_s\left(-2\mathcal{M}(\Re[T^{\Phi}])\right)_{rs} V_s
\end{equation}
For consistency with our publication \cite{Whitfield14a}, we define $K=-2\mathcal{M}\left(\Re[T^{\Phi}]\right)$ as the force-balance operator.  For symmetric matrices, $A=A^{T}$, $\mathcal{M}(A)$ will have the constant vector in its null space.  The gauge freedom physically stems from the irrelevance of the zero of energy. Since we are concerned with time-dependent quantum mechanics, the constant potential only imprints an unobservable global phase on the wave function.  

\section{$V$-representation solver}
In a previous publication \cite{Whitfield14a}, we presented and analyzed an explicit solution for the time-dependent potential necessary for TDDFT provided with the density time-trace of a $V$-representable system.  The algorithm was found to scale polynomially in all input parameters except for the $V$-representability parameter \cite{Whitfield14a} which diverges when the interacting system no longer has a corresponding KS system.  

The algorithm requires, as inputs, the complete time-trace of the density and an consistent initial state $\Phi$ that reproduces the initial density and the initial time derivative of the density.  For numerical implementation, the kinetic energy in the lattice basis is also needed. A explicit solver based on the Runge-Kutta methods~\cite{Dormand80} updates the wave function based on the KS potential computed at each timestep. We discuss some novel aspects of the implementation next.

\subsection{Preparing initial states}
Suppose that an orbital should have an initial time derivative given by $\dot n^{aim}$.
Recall \cite{Whitfield14a} that for a one-electron wave function, $\psi$, the density derivative is given by $\dot n_j=-i \sum_k T_{ki} (\langle a_i^\dag a_k \rangle_\psi-\langle  a_k^\dag  a_i \rangle_\psi)$.  Given that we consider a single electron wave function, the 1-RDM elements can be defined as $\langle a_i^\dag a_j\rangle=\sqrt{n_in_j}e^{i(\phi_j-\phi_i)}$. 

To assign the phases, roughly speaking, we must solve the equation: $\Delta \vec\phi=\frac{\partial\vec\phi}{\partial \dot n} \Delta \dot n$.  This is the content of Newton's method.  We will describe the modifications needed to handle the gauge degree of freedom after deriving the Jacobian, $J_{ij}=\frac{\partial \dot n_i}{\partial\phi_j}$.

Surprisingly, the Jacobian is also given by (minus) the force-balance matrix:
\begin{eqnarray}
\frac{\partial \dot n_i}{\partial \phi_j}
&=&\frac{\partial}{\partial\phi_j}\left(-i \sum_k T_{ki} (\langle a_i^\dag a_k \rangle-\langle  a_k^\dag a _i \rangle)\right)\\
&=&2\sum_k T_{ki} \sqrt{n_kn_i}\; \left[\frac{\partial\sin(\phi_k-\phi_i)}{\partial\phi_j}\right]\\
&=&2\sum_kT_{ki}\sqrt{n_kn_i}\cos(\phi_k-\phi_i)\left[
 \delta_{jk}
-\delta_{ij}
\right]\\
&=& T_{ij}(\langle a_i^\dag a_j\rangle+\langle a_j^\dag a_i\rangle)\nonumber\\
&&-\delta_{ij}\sum_k
T_{jk}(\langle a_k^\dag a_j\rangle+\langle a_j^\dag a_k\rangle)\\
&=&-K_{ij}
\end{eqnarray}

Before applying the Newton method, we must account for the gauge corresponding to the global phase of the wave function. Other manifestations of this gauge degree of freedom are 1) the one-dimensional null space of the Jacobian and 2) the constraint that $\sum \dot n_j=0$.

We can fix the phase of one of the $M$ wave function components in order to fix the gauge.  Suppose the fixed phase is the last, then we only update $\vec\phi_g$ on the $M-1$ remaining components.  The fixed gauge Jacobian $J_g$ is equal to the Jacobian on the first $M-1$ components. 

Putting it all together, the Newton rule for updating the phase vector $\vec \phi$ to a new assignment $\vec \varphi$ is
\begin{equation}
\vec{\varphi}_g=\vec\phi_g -J^{-1}_g (\dot n_g -\dot n_g^{aim})
\end{equation}
Note that dropping the gauge component to get $\dot n_g$ and $\dot n^{aim}_g$ loses no information due to the constraint that $\sum \dot n_j=0$.

Numerical results with a straightforward implementation of Newton's method works quite well provided that the wave function's initial momentum is somewhat close to the target momentum.  This is consistent the with expected performance of Newton's method in other application areas.

As this paper primarily concerns itself with single-electron one-dimensional test cases, we only briefly discuss paths towards adapting the previous method to multi-electron wave functions.  Consider a state with occupied orbitals $\psi_\mu$ for $\mu=1,...,N$.  The total density derivative is merely the sum of each orbital i.e. $\dot n_j=\sum_\mu^N \dot n_j^{(\mu)}$.  Thus, we can apply the Newton method to the first orbital to optimize the phase factors associated with $\psi_1$.  The orthogonality constraints for the remaining $N-1$ orbitals is then enforced by updating the phases of the other orbitals appropriately. 

\subsection{Numerical solutions to the force-balance equation}
To handle the inversion in spite of the non-zero vector in the kernel of the force-balance matrix, we use the truncated singular value decomposition (tSVD).  We also tested Tikhonov regularization but we found that the tSVD works best in the examples tested.  The fixed cut-off used for the truncation corresponds to the maximum allowed $V$-representability parameter.  
	
Note that, at least in the case of a spreading wave function, the additional vectors in the null space are not indicative that the KS system does not exist.  Following the theorem in Ref.~\cite{Farzanehpour12}, one may think that it would at least correspond to the non-uniqueness of the KS. This is trivially true.  

We can understand this non-uniqueness and remedy it easily by considering analogies to the reducibility of Markov chains.  If a Markov chain is reducible, then there is a reordering of the sites such that the Markov matrix can be written in blocks as
\begin{equation}
	P=\left[
		\begin{array}{cc} P_1 &0\\0&P_2
\end{array}
\right].
\end{equation}
Suppose $P_1$ and $P_2$ are both irreducible and aperiodic, such that each has a unique fixed point labeled $\pi_1$ and $\pi_2$ respectively.  Now, $P$ has several fixed points i.e. $(\pi_1,\mathbf{0})^T$, $(\mathbf{0},\pi_2)^T$, and $(\pi_1,\pi_2)^T$.  The interesting state is, of course, $(\pi_1,\pi_2)^T$ while the other two are less interesting.  

In the same way, for densities with disjoint support, relabelling sites will also give $K$ a block structure.  Then the non-trivial solution corresponds to the inhomogeneous solutions in each disjoint region.  So long as each block of $K$ is $V$-representable, the total system remains $V$-representable with a unique non-trivial global solution. When there is a small coupling between two nearly disjoint spaces, this must be handled with some care as perturbation theory is easily applied to eigenvalues but not to eigenspaces. 

\renewcommand{\P}{\mathbb{P}}
We used optimization techniques to tackle this problem \cite{Boyd09}.  Consider the problem of solving $Ax=b$ (with $A=A^T\in \mathbf{M}(\mathcal{R})$ and $x,b\in \mathcal{R}$) on subspace defined by $\P$ s.t. $\P^2=\P$ 	
\begin{eqnarray}
		r&=& |A\P x-b|^2\\
		&=&\left|\left(
		\begin{array}{cc}
			A_{1}&A_{21}\\A_{12}&A_{22}
		\end{array}
		\right)\left(
		\begin{array}{c}
			x\\0
		\end{array}
		\right)-\left(
		\begin{array}{cc}
			b_1\\b_2
		\end{array}
		\right)\right|^2\\
		&=&\phantom{+} x^T A_{1}^2x-x^T  A_{1} b_1-b_1^T A_{1}x+b_1^2+b_2^2\nonumber\\
		&&+x^T A_{12}A_{21}x-x^T  A_{12} b_2-b_2^T A_{21}x
\end{eqnarray}
Setting the derivative of $r$ with respect to the vector $x$ to zero implies 
\begin{equation}
		x=(A_{1}^2 +A_{12} A_{21})^{-1}(A_{1} b_1+A_{12}b_2).
\end{equation}

In our numerical implementation, we found that this expression can be improved by noting that $A_{21}$ is a perturbation to the matrix $A_1$.  We rearranged the expression to achieve better numerical results:
	\begin{equation}
		x= (A_{1} +A_1^{-1}A_{12} A_{21})^{-1}( b_1+A_{1}^{-1}A_{12}b_2)
		\label{eq:x}
	\end{equation}
The cost of the numerical inversion depends on the size of $A_1$ since two matrix inversions are needed to evaluate \eqref{eq:x} whereas the full inversion only requires one matrix inversion.  In numerical experiments, we found that when dim$[A_1]$ is less than 60\% of the full space, \eqref{eq:x} is faster.

Additional improvements were made by employing a double truncation technique whereby one domain is defined by region where the density is non-trivial and a second domain is defined as the region where $\ddot n$ is non-trivial.  In the examples where the wave function is spreading, the region corresponding to $\ddot n$ is larger.  If we let $\P'$ be the larger of the two domains and $\P$ the smaller, the procedure employed  solves for the KS potential within region $\P'$ and then truncates the potential to region $\P$.

Another issue that affects numerical performance is the choice of the gauge.  The two choices we considered were 1) fixing the gauge such that the mid-point of the domain defines zero potential, 2) fixing the gauge as the mean of the potential.  The first choice is natural in many of the examples since the potential is known to be zero at the mid-point.  This is not generically the case so the second choice offers a sensible alternative. By choosing the gauge based on the average value of the potential, the norm of the Hamiltonian takes its minimal value over all gauge choices. This improves numerical stability when the timesteps are based on the norm of the Hamiltonian but we also found that it can cause erratic jumps in the KS potential due to numerical noise at the boundaries. 

\section{Numerical examples}
In this study, we tested our solver on some simple numerical examples to validate the solver: ballistic spreading, a superposition of particle-in-a-box states and the ``discovery'' of a constant potential.

The first example, ballistic spreading of the wave function, served as a simple test of the proposed truncation procedure and provided insights into the best methods for handling spreading.  It was used to compare several different truncation procedures and served as validation for the procedure presented here.  The numerical noise led to a potential norm on the order of $10^{-8}$ as the wave function spread ballistically.  The second example merely confirmed the results of \cite{Li08} and also served as a test case for the procedure when the density is evolving in a confined space.  

\begin{figure}[t]
	\begin{center}
	\includegraphics[width=\columnwidth]{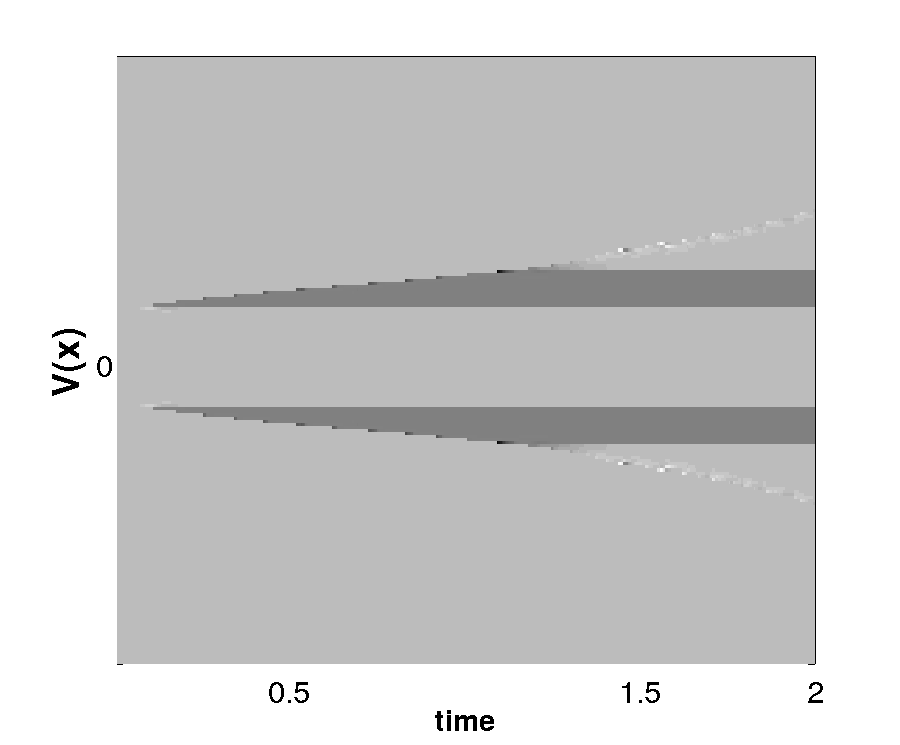}
	\end{center}
	\caption{As the density spreads, it discovers the potential and this is reflected in the solutions to the $V$-representation problem depicted here. The numerical errors at the boundary of the wave packet's extent are all less than unity and the average maximum error at the boundary is 0.3195.} 
	\label{fig:1}
\end{figure}

\begin{figure}[t]
	\begin{center}
	\includegraphics[width=\columnwidth]{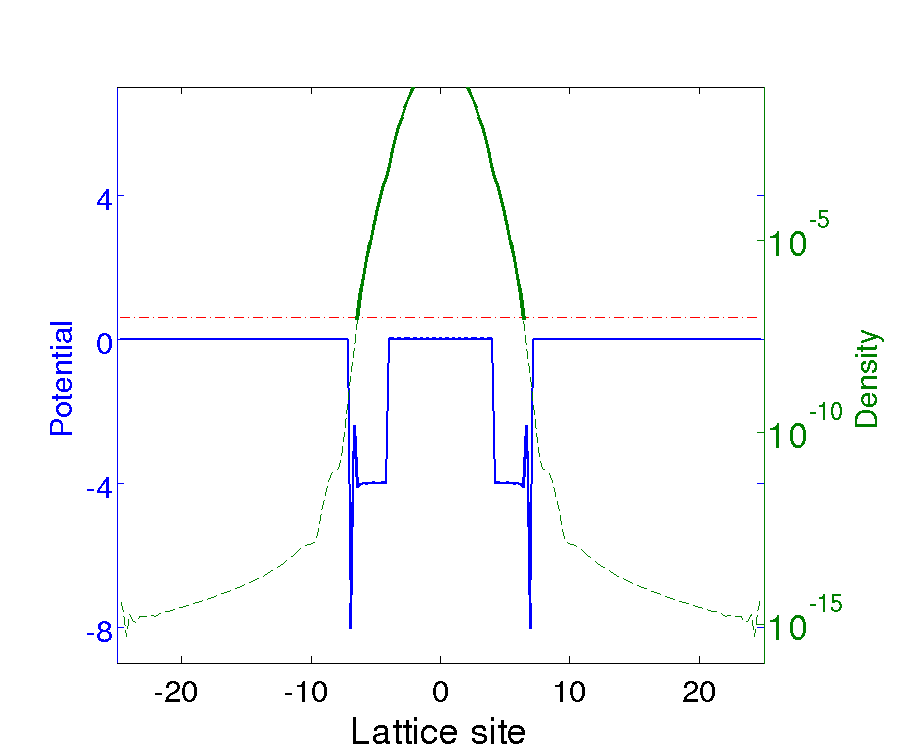}
	\end{center}
	\caption{The difficulty of solving the $V$-representation problem is illustrated by the potential and density at time $1.0671$ as plotted here.  The red dot-dashed line indicates the cutoff used for the density and the heavy blue line represents the potential.} 
	\label{fig:2}
\end{figure}

The last test case considered is a density which evolves from an initial Gaussian distribution with unit variance centered at the origin and zero momentum in a fixed potential defined by
\begin{equation}
	V(x)=\left\{ \begin{array}{cc}
		-4 &|x-5.5|<1.5\\
		-4 &|5.5-x|<1.5\\
		0 & \textrm{otherwise}
	\end{array}
		\right.
\end{equation}
When the propagation starts, the density does not know about the potential wells to the right and left, but as it evolves the second derivative and the density itself, learn about the potential wells.  The numerical procedure remains stable throughout as depicted in Fig.~\ref{fig:1}. The discovery of the potential is clearly illustrated as the density begins to evolves into the lower potential regions. The worst numerical error encountered during the numerical procedure occurs when the density reaches the outer edges of the potential wells as shown in Fig.~\ref{fig:2}. This can be understood as the solver's inability to decide what the potential should be beyond the area that the density has seen.  

Here, it is interesting to note that the spreading of the density only occurs at second order. From $\partial_t \hat n=i[\hat H,\hat n]$
\begin{equation}
\partial_t \hat n_j =-i \sum_k T_{kj}( a_j^\dag  a_k- a_k^\dag  a_j)
\end{equation}
Suppose that $\ket{\Psi}=\hat\Psi\ket{\Omega}=\sum C_K \ket{K}$ with $\hat K=\prod_{i=1}^M \left(a_i^\dag\right)^{K_i}$ has no support on site $j$. 
Then $[ a_j,\hat \Psi]\ket{\Omega}=0$ since $\hat \Psi$ has no support on $j$. 
Therefore, $\bra{\Omega} \hat \Psi^\dag a_k^\dag  a_j \hat \Psi \ket{\Omega}$ is zero as well.  Hence, 
\begin{equation}
	\bra{\Psi} \partial_t \hat n_j\ket{\Psi}= 
	-i\sum_k T_{kj}\left(\bra{\Omega} \hat \Psi^\dag  a_j^\dag a_k \hat\Psi\ket{\Omega}-\textrm{h.c.}\right)=0
\end{equation}
Thus, density spreads to new regions only at second order in time.

\section{$V$-representability}
We now make a few comments on the existence of solutions to the $V$-representation problem.
A previous theorem \cite{Farzanehpour12} showed that ground states many-body interacting systems are always $V$-representable in the neighborhood of the initial time.  They showed that the matrix $K$ has only one zero eigenvalues and is positive definite in the space of inhomogeneous potentials. It should be noted that the theorem does not characterize the $V$-representability parameter thus numerical stablility is not ensured.

Here we use simpler arguments to provide additional characterizations of the spectrum of $K$ in the single-electron case. This coincides with the previous theorem for the ground state but generalizes to all eigenstates.

\textbf{Theorem:} Given non-degenerate $\psi$ such that $H\psi=E_k\psi$ for $H=T+V\in \mathbf{M}(\mathcal{R})$, $K(\psi)$ has $k-1$ negative eigenvalues and $M-k+2$ positive eigenvalues.

\textbf{Proof:}
Assuming that $\psi$ is an eigenstate of $H=T+V$ with eigenvalue $\lambda$, then $H^{(\lambda)}=T-(\lambda \mathbf1 -V)=T-D$ has $\psi$ in its null space. Rearranging, $(T+D)\psi=0$ implies $\sum_k T_{jk}\psi_k=D_j\psi_j$. Before using the definition of the force-balance equation, it is important to note that eigenvectors of symmetric matrices are real. Hence, for a single-particle in the eigenstate $\psi$:
\begin{eqnarray}
K_{ij}&=&-2T_{ij}\psi_i\psi_j +2\delta_{ij} \psi_j \left(\sum_k T_{jk}\psi_k\right)\\
&=&-2T_{ij}\psi_i\psi_j-2\delta_{ij}D_j\psi_j^2\\
&=&\sum_{mn}(\delta_{mi}\psi_i)2(T_{mn}-\delta_{mn}D_n)(\delta_{nj}\psi_j)\\
\end{eqnarray}
Since $\psi$ has full support, $S_{mj}=\delta_{mj}\psi_j$ is non-singular and the number of $(+/0/-)$ eigenvalues are the same for $H^{(\lambda)}$ and $K(\psi)$ by Sylvester's theorem~\cite{Horn05}.
$\square$

According to numerical tests, if $\psi$ does not have full support then $K$ will have an additional vector in the null space.  This is consistent with the theorems from Ref.~\cite{Farzanehpour12}. The interacting extension of the present theorem does not seem to hold although we found that the many-body non-interacting ground state gives rise to $K$ with the same inertia as $H^{(\lambda_0)}$ consistent with the previous findings \cite{Farzanehpour12}.

\section{Outlook}

Next steps for the long term project begun here are 
the study of interacting electronic examples,
comparisons and combinations with the implicit fix-point methods~\cite{Ruggenthaler11,Ruggenthaler12,Nielsen13}, and
designing algorithms for building multi-particle initial states.
Theoretical questions to be tackled include understanding $V$-representability of open system evolutions, continuum limits and intersections with quantum computing. Previously, we have shown that the $V$-representation problem can be solved efficiently using a quantum computer but it still remains an open question if all quantum computations remains efficiently simulatable with TDDFT given access to efficient solutions to the $V$-representation problem.

\acknowledgments
I thank the Vienna Center for Quantum Science and Technology and the Ford Foundation for financial support.


\begin{thebibliography}{10}
\expandafter\ifx\csname url\endcsname\relax\def\url#1{\texttt{#1}}\fi

\bibitem{Mazziotti11}
\Name{Mazziotti D.~A.} \REVIEW{Chem. Rev.}{112}{2012}{244}.

\bibitem{Coulson60}
\Name{Coulson C.~A.} \REVIEW{Rev. Mod. Phys.}{32}{1960}{170}.

\bibitem{Lui07}
\Name{Lui Y.-K., Christandl M. \and Verstraete F.} \REVIEW{Phys. Rev.
  Lett.}{98}{2007}{110503}.

\bibitem{Lackner15}
\Name{Lackner F., B\ifmmode~\check{r}\else \v{r}\fi{}ezinov\'a I., Sato T.,
  Ishikawa K.~L. \and Burgd\"orfer J.} \REVIEW{Phys. Rev. A}{91}{2015}{023412}.

\bibitem{Hohenberg64}
\Name{Hohenberg P. \and Kohn W.} \REVIEW{Phys Rev}{136}{1964}{B864}.

\bibitem{Schuch09}
\Name{Schuch N. \and Verstraete F.} \REVIEW{Nature Physics}{5}{2009}{732}.

\bibitem{Whitfield14a}
\Name{Whitfield J.~D., Yung M.-H., Tempel D.~G., Boixo S. \and Aspuru-Guzik A.}
  \REVIEW{New J. Phys.}{16}{2014}{083035}.

\bibitem{Tempel12}
\Name{Tempel D.~G. \and Aspuru-Guzik A.} \REVIEW{Sci. Rep.}{2}{2012}{391}.

\bibitem{Leeuwen99}
\Name{van Leeuwen R.} \REVIEW{Phys. Rev. Lett.}{82}{1999}{3863}.

\bibitem{Baer08}
\Name{Baer R.} \REVIEW{J. Chem. Phys.}{128}{2008}{044103}.

\bibitem{Li08}
\Name{Li Y. \and Ullrich C.~A.} \REVIEW{J. Chem. Phys.}{129}{2008}{044105}.

\bibitem{Farzanehpour12}
\Name{Farzanehpour M. \and Tokatly I.~V.} \REVIEW{Phys. Rev.
  B}{86}{2012}{125130}.

\bibitem{Ruggenthaler11}
\Name{Ruggenthaler M. \and van Leeuwen R.} \REVIEW{Europhys.
  Lett.}{95}{2011}{13001}.

\bibitem{Ruggenthaler12}
\Name{Ruggenthaler M., Giesbertz K. J.~H., Penz M. \and van Leeuwen R.}
  \REVIEW{Phys. Rev. A}{85}{2012}{052504}.

\bibitem{Nielsen13}
\Name{Nielsen S. E.~B., Ruggenthaler M. \and van Leeuwen R.} \REVIEW{Europhys.
  Lett.}{101}{2013}{33001}.

\bibitem{Dormand80}
\Name{Dormand J.~R. \and Prince P.~J.} \REVIEW{J. Comp. Appl.
  Math.}{6}{1980}{19}.

\bibitem{Boyd09}
\Name{Boyd S. \and Vandenberghe L.} \Book{Convex Optimization} (Cambridge
  University Press) 2009.

\bibitem{Horn05}
\Name{Horn R.~A. \and Johnson C.~R.} \Book{Matrix analysis} (Cambridge
  University Press) 2005.

\end{thebibliography}
\end{document}